\documentclass[aps,preprintnumbers,superscriptaddress,
              showpacs,nofootinbib]{revtex4}
 \newcommand{\PRE}[1]{}       
\usepackage{amsmath,amsfonts,epsfig,graphicx}
\usepackage{multirow}
\usepackage[latin1]{inputenc}
 \usepackage{amssymb}
 \usepackage{verbatim}
\usepackage{float}
\usepackage{slashed} 

\newcommand{\MGMCatNLO}{MadGraph5\_aMC@NLO} 
\newcommand{\madgraph}{{\sc MadGraph}}

\newcommand{\pythia}{{\sc Pythia}}

\newcommand{\delphes}{{\sc Delphes}}

\begin{document}
\title{Polarization fraction measurement in ZZ scattering using deep learning}

\author{Junho Lee}
\affiliation{
Department of Physics and State Key Laboratory of Nuclear Physics and Technology, Peking University, Beijing, 100871, China}

\author{Nicolas Chanon}
\affiliation{
Institut de Physique Nucl\'{e}aire de Lyon, Universit\'{e} de Lyon, Universit\'{e} Claude Bernard Lyon 1, CNRS-IN2P3, Villeurbanne 69622, France }

\author{Andrew Levin}
\affiliation{
Department of Physics and State Key Laboratory of Nuclear Physics and Technology, Peking University, Beijing, 100871, China}

\author{Jing Li}
\affiliation{
Department of Physics and State Key Laboratory of Nuclear Physics and Technology, Peking University, Beijing, 100871, China}

\author{Meng Lu}
\affiliation{
Department of Physics and State Key Laboratory of Nuclear Physics and Technology, Peking University, Beijing, 100871, China}

\author{Qiang Li}
\affiliation{
Department of Physics and State Key Laboratory of Nuclear Physics and Technology, Peking University, Beijing, 100871, China}

\author{Yajun Mao}
\affiliation{
Department of Physics and State Key Laboratory of Nuclear Physics and Technology, Peking University, Beijing, 100871, China}


\begin{abstract}
Measuring longitudinally polarized vector boson scattering in the ZZ channel is a promising way to investigate unitarity restoration with the Higgs mechanism and to search for possible new physics. We investigated several deep neural network structures and compared their ability to improve the measurement of the longitudinal fraction $Z_L Z_L$. 
Using fast simulation with the Delphes framework, a clear improvement is found using a previously investigated `particle-based' deep neural network on a preprocessed dataset and applying principle component analysis to the outputs.
A significance of around 1.7 standard deviations can be achieved with the integrated luminosity of 3000 $fb^{-1}$ that will be recorded at the High-Luminosity LHC. The technique developed in this article is also useful
to other LHC analyses involving helicity fraction measurement.
\end{abstract}

\keywords{Vector Boson Scattering, Polarization, LHC, Deep Neural Network}
\pacs{HL-LHC, vector boson scatter, electroweak}

\maketitle

Vector boson scattering (VBS) is a rare standard model (SM) process which plays a crucial role in electroweak symmetry breaking.
The LHC and High-Luminosity LHC (HL-LHC) have enormous potential to both initially observe and study the features of rare processes such as VBS.
Our knowledge of the VBS topology at hadron colliders can be combined with advanced data analysis techniques, such as deep learning, to make this pursuit even more promising.

Many VBS studies have been performed based on LHC data, including measurements of  W$^{\pm}$W$^{\pm}$~\cite{Sirunyan:2017ret, Aaboud:2019nmv}, W$^{\pm}$Z~\cite{Sirunyan:2019ksz, ATLAS:2018ucv} and $Z\gamma$~\cite{Khachatryan:2017jub,CMS:2019iuv}.
The topic of this paper is the channel ZZ~$\rightarrow$ 4l.
While this channel has the advantage of a clean final state, it suffers from a low production cross section, small branching-ratio of the Z boson to charged leptons, and two large irreducible QCD backgrounds, the production of ZZ via quark-antiquark annihiliation (qqZZ) and via a gluon box diagram (ggZZ).
ZZ scattering has recently been observed by ATLAS with a significance larger than 5 standard deviations, using 139 $fb^{-1}$ of LHC Run II data collected at $\sqrt{s}=13$~TeV~\cite{ATLAS-CONF-2019-033}.
A prior measurement made by CMS with $35.9fb^{-1}$ collected at $\sqrt{s}=13$~TeV reported an observed significance of 2.7 standard deviations ~\cite{Sirunyan:2017fvv}.

\begin{figure}[htbp]
  \begin{center}
  \includegraphics[width=0.4\textwidth]{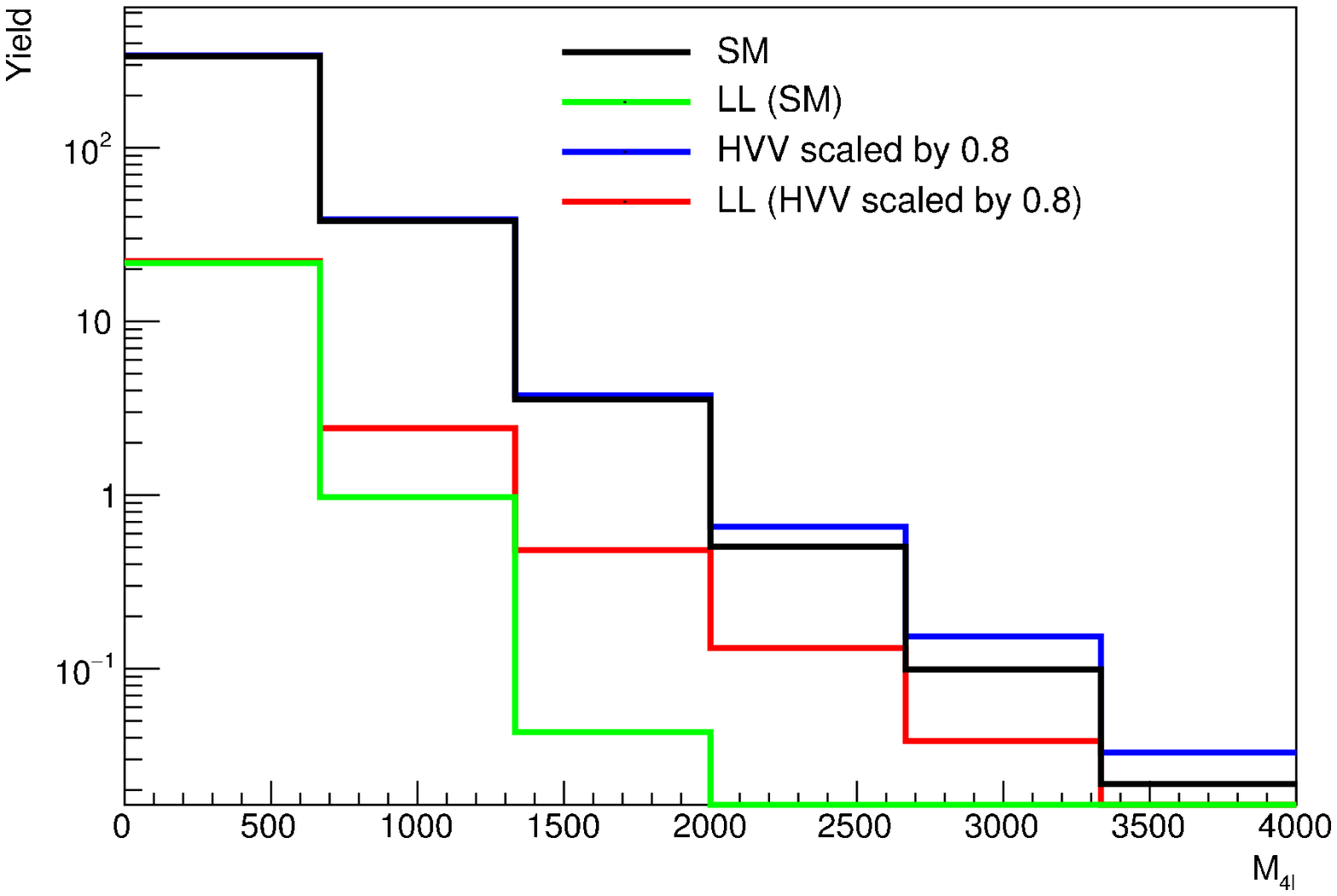}
  \includegraphics[width=0.4\textwidth]{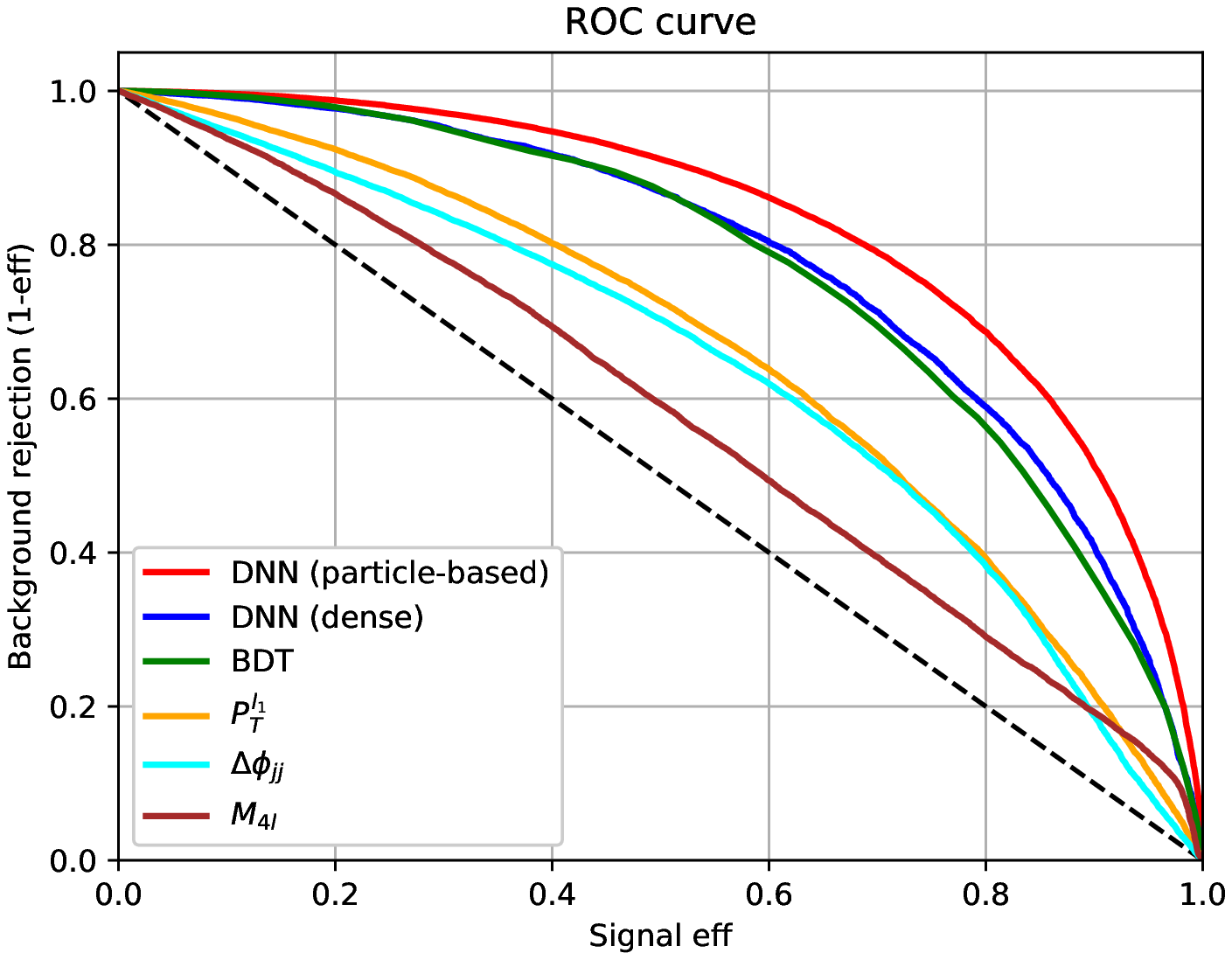}
   \caption{[left:] Comparision of 4 lepton invariant mass distributions, normalized to the HL-LHC scenario of 3000 $fb^{-1}$, among the SM and its LL component, the SM but with Higgs and W or Z boson couplings scaled by a factor of 0.8, and its LL component. [right:] Comparison of the discriminating power of a particle-based DNN, a dense DNN, a BDT, and several kinematic discriminants. The signal is LL and the background is TT and TL.
    }
    \label{fig:ROC_pilot_curves}
  \end{center}
\end{figure}

Measuring the longitudinally polarized component of VBS (the LL component) is a critical next step for the field, as it is closely related to the important theoretical property of unitarity restoration, through Higgs and possible new physic~\cite{Chang:2013aya,Lee:2018fxj}. In the context of Higgs and VBS discoveries, it becomes one of the next big targets, to test directly that Higgs regulates Higgs Goldstone scattering~\cite{Quigg:2018llo}. Prbing LL scattering can also provide alternative and model independent way of measuring Higgs coupling~\cite{Brooijmans:2014eja}. Fig.~\ref{fig:ROC_pilot_curves} [left] shows comparision of 4 lepton invariant mass distributions among the SM and its LL component, the SM but with Higgs and W or Z boson couplings scaled by a factor of 0.8, and its LL component. In this example, the LL component can be sensitive to Higgs couplings with massive gauge bosons, as any deviations from the SM prediction leads to large enhancements of the LL mode especially at high mass tail. However, because the LL cross-section is only $\backsim$10\% of the sum of the transverse (TT) and mixed (TL) cross-sections, and has very similar features, this measurement will be very challenging, and advanced techniques will be essential for its success, as can be seen in Fig.~\ref{fig:ROC_pilot_curves} [right] where large improvement can be achieved.

Previous studies include WW scattering studies which employ a two-dimensional fit and a deep neural network (DNN)~\cite{CMS:2018zxa, Lee:2018xtt} and ZZ scattering studies that employ a boosted-decision tree (BDT)~\cite{CMS:2018mbt}.
However, due to a low signal yield expected in the ZZ case, a significance of 1.4$\sigma$ is expected for LL with 3000 $fb^{-1}$ collected at $\sqrt{s}=14$~TeV. In this study, we compare the performance of several machine-learning models, including a BDT~\cite{Roe:2004na} as implemented in TMVA~\cite{TMVA}, and a DNN as implemented in the Keras library~\cite{Keras} with Tensorflow back-end~\cite{Tensorflow}.
There has been a performance comparison study for extracting LL component on same-sign WW channel, using particle-based DNN, dense DNN and BDT~\cite{Lee:2018xtt}. The result shows that the particle-based DNN performs the best. 
We further enhanced the particle-based DNN classification power by applying standardization and Yeo-Johnson power transformation~\cite{10.1093/biomet/87.4.954} to each input variable.
Finally, principle component analysis (PCA) is applied to the outputs of the DNN, and then two or three dimensional fits are performed, to achieve further enhancement of the signal significance.

VBS ZZ and QCD ZZ background samples are simulated using {\MGMCatNLO}~\cite{Alwall:2014hca} for the hard process and \pythia\,6~\cite{Sjostrand:2003wg} for the parton showering and hadronization.
\delphes~version 3~\cite{deFavereau:2013fsa} was used for detector simulation, and was configured to simulate the CMS HL-LHC detector (`pileup' has been neglected as in ref.~\cite{Searcy:2015apa}).
The LL, TL, and TT samples are obtained from the VBS ZZ sample and processed separately starting from generator level.
Z bosons of each sample are decayed into charged lepton pairs by \madgraph's DECAY package, while keeping the polarization information for each of the boson. 
The following event selection is applied to reconstructed Delphes objects. We require 4 leptons which can be clustered into two pairs of Z boson candidates, with each Z boson candidate consisting of two oppositely-charged same-flavor leptons and an invariant mass ($m_{\text{ll}}$) satisfying $60~{\rm GeV} < m_{\text{ll}} < 120~{\rm GeV}$.
If there is more than one scenario of ZZ combination, we select the minimum $(m_{\text{ll}_1} - 90.188)^2 + (m_{\text{ll}_2} - 90.188)^2$ case .
Each of the selected leptons' transverse momentum ($p_T^l$) must be larger than $5~{\rm GeV}$. The leading and sub-leading leptons' transverse momentum must be larger than $20~{\rm GeV}$ and $10~{\rm GeV}$, respectively.
We select jets that have transverse momentum ($p_T^j$) larger than $25~{\rm GeV}$ and an absolute value of pseudorapidity($|\eta_{\text{j}}|$) smaller than 4.7. And we require at least two selected jets.
In order to suppress QCD-induced backgrounds, we require that the invariant mass of the two leading jets ($m_{\text{jj}}$) be larger than $400~{\rm GeV}$, that the absolute value of their pseudorapidity separation ($|\Delta\eta_{\text{jj}}|$) be larger than 2.4, and that the event contain no b-tagged jets.
After event selection, 100,000, 150,000, 240,000, 48,000, and 40,000 unweighted events are left for LL, TL, TT, qqZZ and ggZZ, respectively. 

We now compare the LL vs. LT and TT discriminating power of several variables including machine learning discriminants and kinematic variables.
We consider two DNNs, a BDT, transverse momentum of leading lepton ($P_T^{l_1}$), and azimuthal angle difference between the two leading jets ($|\Delta\phi_{\text{jj}}|$).
The DNNs and the BDT models take as input the four momenta of four leptons and two jets.
We have constructed two differently structured DNN models, a dense DNN and a particle-based DNN, a structure previously explored in ref~\cite{Lee:2018xtt}. 
The dense DNN model uses 6 hidden layer with 150 hidden units. In the particle-based model, four momenta of each particle makes one input layer. 
The model contains two hidden layers with 10 nodes for each particle. Leptons and jets are merged with 3 layers of 20 nodes before they are merged into one big layer of 60 nodes. Finally, 2 layers of 100 nodes are added ahead of the final output layer.
Both DNN models implement a `relu' activation function on every hidden unit, while a sigmoid function is used at each output node. Also, He's uniform~\cite{DBLP:journals/corr/HeZR015} function is adopted for weight initialization, and the `adam' optimizer ~\cite{2014arXiv1412.6980K} with a learning rate of 0.001 is used to train each DNN.
No overtraining was found -- the loss values are comparable for the training and test samples, as is the distribution of the DNN output.
Fig.~\ref{fig:ROC_pilot_curves} [right] compares the performance of each discriminant, where LL stands for as signal and TL and TT are taken as a background. 
The particle-based DNN model has the best discrimination power among all of the discriminants and will be the main focus of this paper.

\begin{figure}[htbp]
  \begin{center}
  \includegraphics[width=0.4\textwidth]{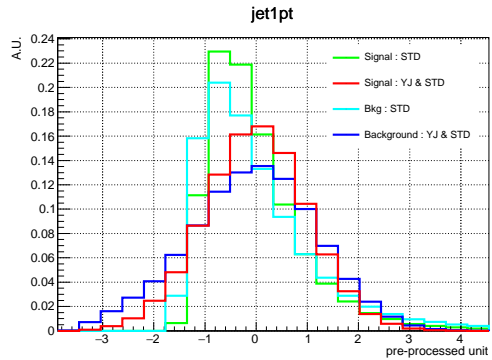}
    \caption{Comparison of different data preprocessing methods. The signal is LL and the background is TL, TT, qqZZ, and ggZZ. The data preprocessing is applied to both the signal and the background.}
    \label{fig:YJ_jet1pt}
  \end{center}
\end{figure}

\begin{figure}[htbp]
  \begin{center}
  \includegraphics[width=0.4\textwidth]{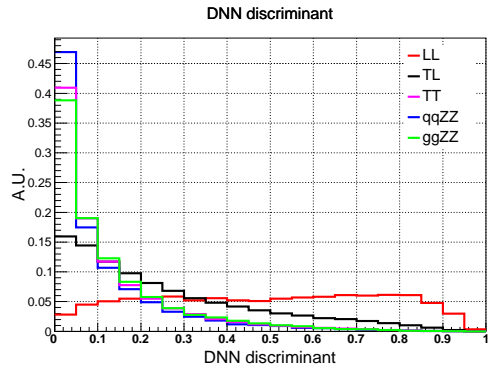}
    \caption{Distribution of the LL score of the particle-based DNN for each sample. YJ\&STD data preprocessing was applied to each input variable.}
    \label{fig:DNN_dist_LL}
  \end{center}
\end{figure}

\begin{figure}[htbp]
  \begin{center}
  \includegraphics[width=0.4\textwidth]{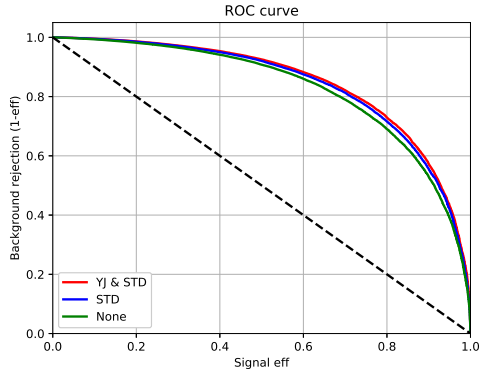}
    \caption{Comparison of data preprocessing methods on the discriminating power of the particle-based DNN. The LL score of each DNN output has been used to make the ROC curves.}
    \label{fig:ROC_dataPreProcess}
  \end{center}
\end{figure}

Next, we re-organize the particle-based DNN output structure such that it has five output nodes instead of two output nodes, with the five output nodes corresponding to LL, TL, TT, qqZZ, and ggZZ scores, respectively.
We also studied the impact of several data pre-processing methods on the DNN performance, including standardization (STD), Yeo-Johnson power transformation~\cite{10.1093/biomet/87.4.954} together with standardization (YJ\&STD), and no preprocessing.
The STD uses the LL train dataset as a base dataset to construct a transformation which transforms each variable distribution such that its mean and standard deviation are 0 and 1.
The standardization of the base dataset is applied to all of the remaining training and test datasets for the signal and all of the backgrounds. Yeo-Johnson power transformation makes the distribution of the LL train dataset a normal distribution. The transformation is an extension of Box-Cox power transformation~\cite{10.2307/2984418} onto negative values, and thus is well suited to our input variables.
Fig.~\ref{fig:YJ_jet1pt} shows a comparison of the distribution of $P_T^{j_1}$ when different data preprocessing methods are applied, where the signal is LL and the background is TL, TT, qqZZ and ggZZ merged together.
The LL distribution for the DNN output in the YJ\&STD case is shown in Fig.~\ref{fig:DNN_dist_LL}.
Fig.~\ref{fig:ROC_dataPreProcess} shows the impact of the different preprocessing methods on the ROC curves for the particle-based DNN. This result shows that the best discriminating power comes from YJ\&STD. We thus adopt this preprocessing method.

The signal significance is calculated based on the Asimov dataset~\cite{Cowan2011}. 
We report a significance for three machine learning methods: 1) a two-step BDT~\cite{CMS:2018mbt}. 2) a two-step DNN, and 3) a DNN with PCA applied to the outputs (DNN-PCA), which are described in detail below.

1) The two-step BDT first classifies QCD vs. VBS (BDT1) and then classifies between LL and all backgrounds (BDT2).
BDT2 was trained on the events left after a selection on the output of BDT1, which maximizes $S/\sqrt{B}$.
We obtain a significance of 1.41$\sigma$ when only statistical uncertainties are considered, and a significance of 1.23$\sigma$ when a 10\% uncertainty is applied both on signal and background in addition to statistical uncertainties. 
These numbers are comparable to the expected significance, 1.4$\sigma$, reported in ref.~\cite{CMS:2018mbt}, which considers statistical uncertainties, systematic uncertainties, and 10\% uncertainty in the ggZZ background yield.


2) Similarly to the two-step BDT, in the two-step DNN the second DNN is applied on the output of the first DNN. 
The first DNN structure is the previously described particle-based DNN, while a shallow dense NN is used as second DNN.
Since there are only five output nodes on the first DNN, which represents LL, TL, TT, qqZZ, and ggZZ score, the second DNN is shallow with 2 hidden layers. Unlike the two-step BDT, no cut is applied on any part of two-step DNN. 
The expected significance obtained with this method is 1.47$\sigma$ (1.38$\sigma$ when systematic uncertainties are considered)
The significances obtained with other data preprocessing methods are listed in Table~\ref{tab:significance_1D}.

\begin{figure}[htbp]
  \begin{center}
  \includegraphics[width=0.4\textwidth]{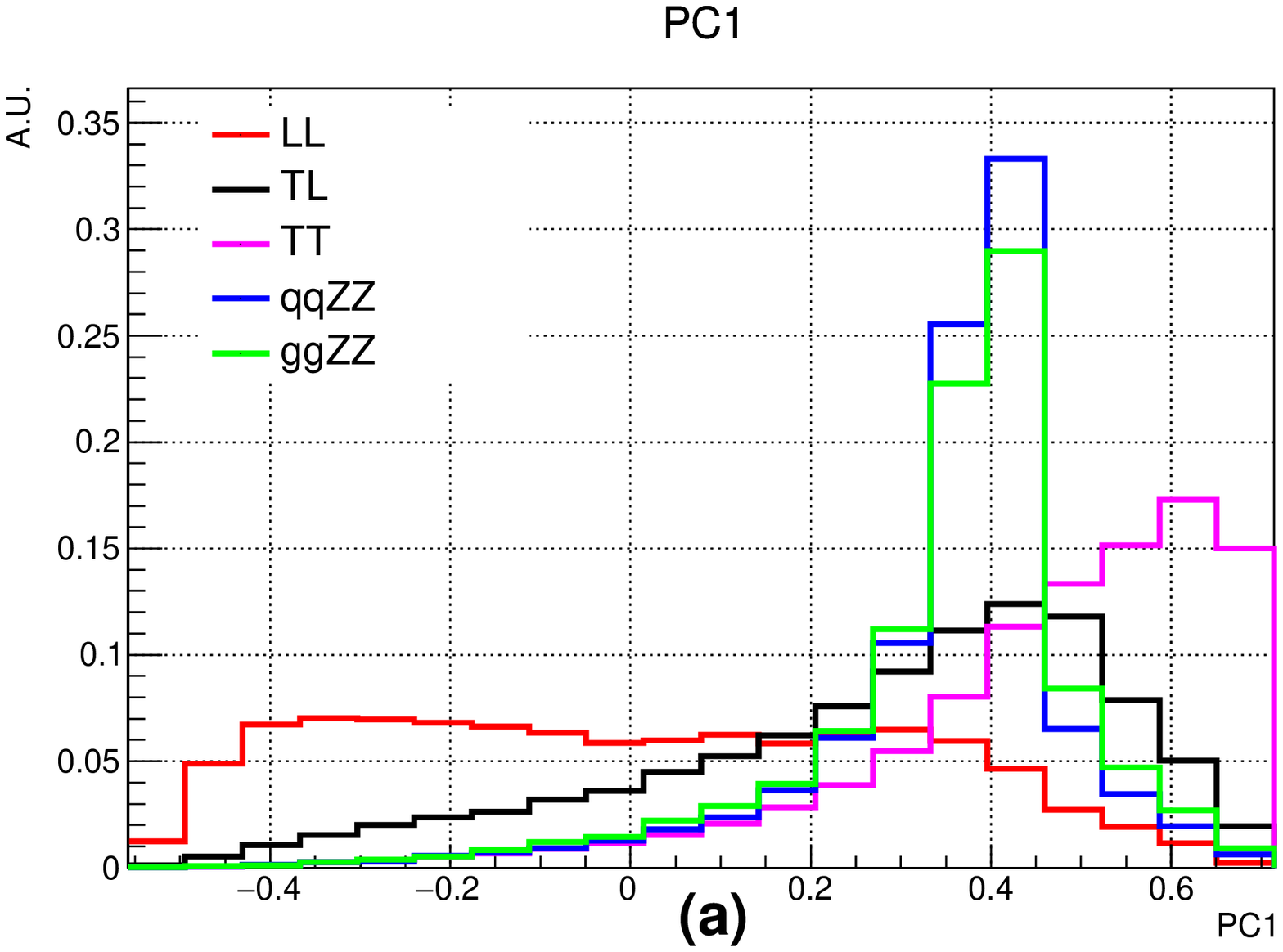}
  \includegraphics[width=0.4\textwidth]{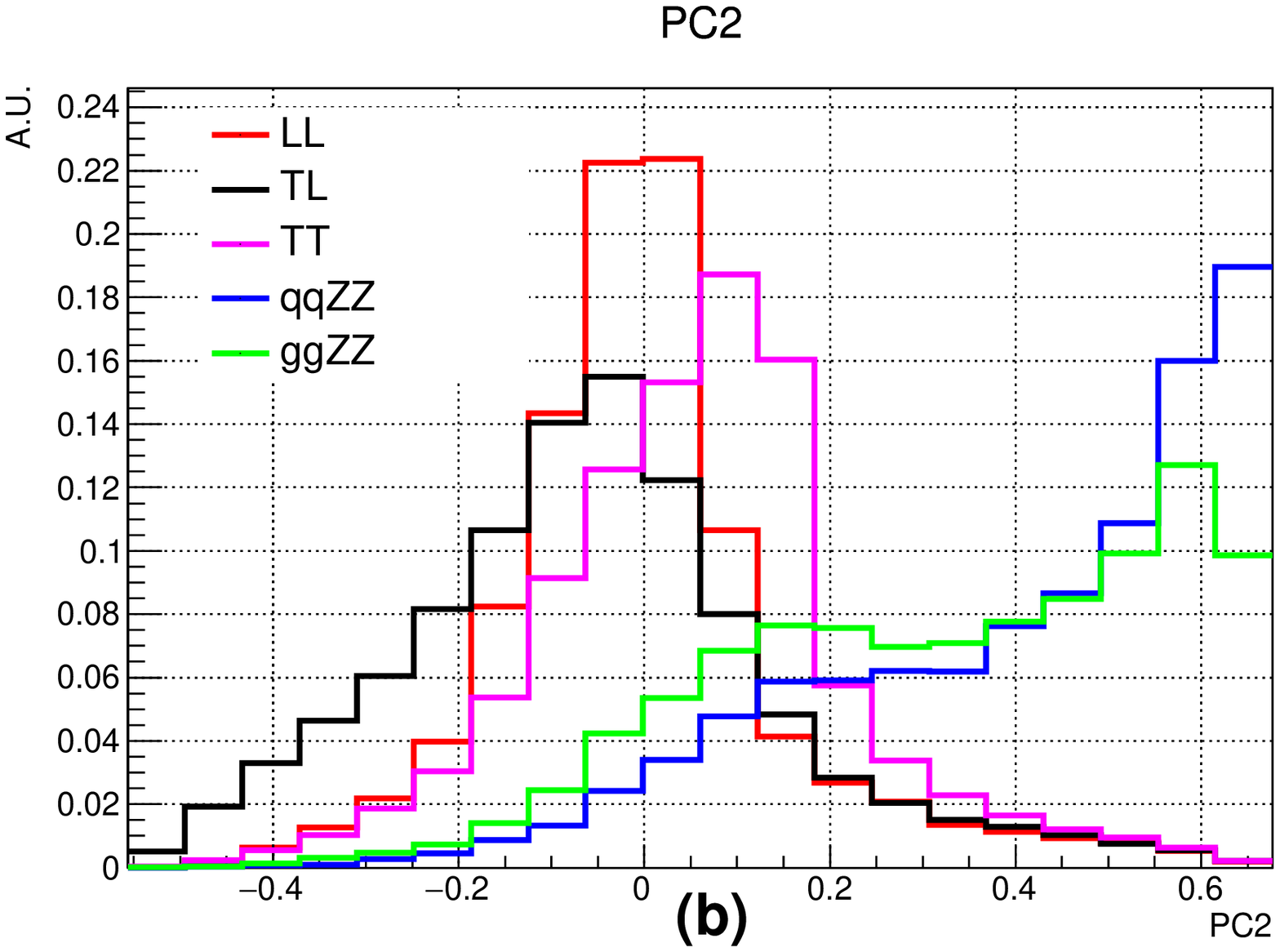}
  \includegraphics[width=0.4\textwidth]{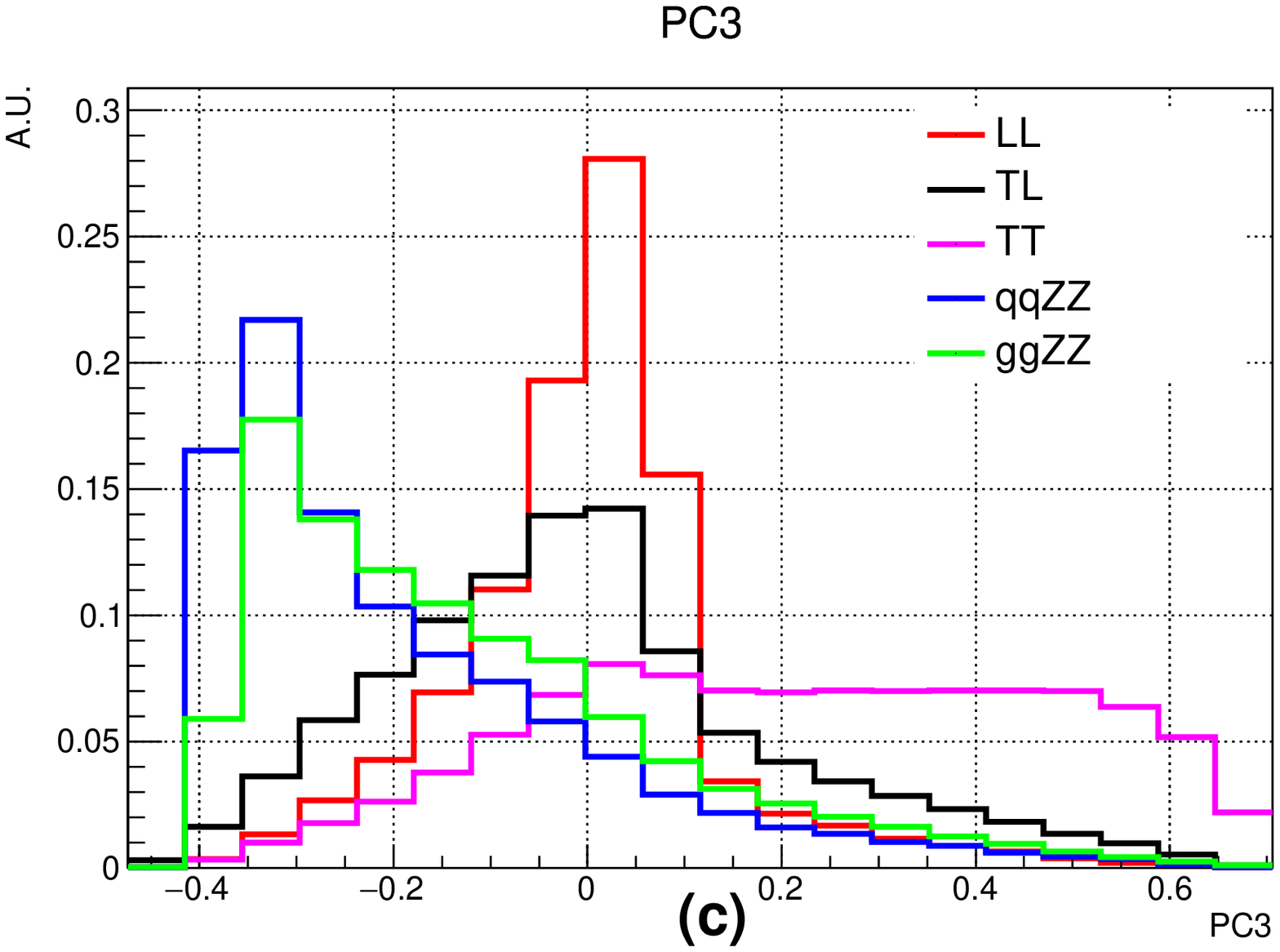}
    \caption{Leading three principle component distribution for each sample.}
    \label{fig:PCA_dists}
  \end{center}
\end{figure}

3) The DNN-PCA has only one particle-based DNN, the same particle-based model as was used in the first part of the two-step DNN, but with principle component analysis (PCA) implemented on its output.
The PCA algorithm rotates the original axis of the features into a new axis containing decorrelated  features.
More specifically, given n-dimension target-data distribution, the first principal component has the largest possible variance and each succeeding component in turn has the highest variance possible under the constraint that it is orthogonal to the preceding components~\cite{wiki:xxx}. 
The five dimensional output from the particle-based DNN is transformed into five principle components. Distributions of the leading three principle components for each sample is shown in Fig.~\ref{fig:PCA_dists}. Additionally, the explained variance ratio of the PCs is listed in Table~\ref{tab:pca_var}. Using PC1 as a discriminant, the signal significance is found to be 1.55$\sigma$ (1.46$\sigma$ including systematic uncertainty).

\begin{table}[htbp]
\begin{center}
{\footnotesize
\begin{tabular}{|c|c|c|c|c|c|}
  \hline 
 Uncertainty & BDT & DNN &STD DNN & YJ\&STD DNN& DNN-PC1\\
 \hline
 Statistical& 1.41$\sigma$ & 1.42$\sigma$ & 1.43$\sigma$ & 1.47$\sigma$ & 1.55$\sigma$\\
 Stat. \& syst. & 1.23$\sigma$ & 1.31$\sigma$ &1.33$\sigma$ & 1.38$\sigma$ & 1.46$\sigma$\\
  \hline
\end{tabular}
}
\end{center}
\caption{Expected signal significance based on different algorithms. The `Statistics' row stands for statistical uncertainty only, while `Stat. \& syst.' includes both statistical and systematic uncertainties.}
\label{tab:significance_1D}
\end{table}

\begin{table}[htbp]
\begin{center}
{\footnotesize
\begin{tabular}{|c|c|c|c|c|c|}
  \hline 
 Principle component & PC1 & PC2 & PC3  & PC4 & PC5\\
 \hline
 Explained variance ratio & 64.8\% & 18.1\% & 13.0\% & 4.2\% & $<$~0.1\% \\
  \hline
\end{tabular}
}
\end{center}
\caption{Explained variance ratio of each principle component. The first, second, and third leading PCs' explained variance ratio exceed 10\%.}
\label{tab:pca_var}
\end{table}
  
\begin{figure}[H]
  \begin{center}
  \includegraphics[width=0.4\textwidth]{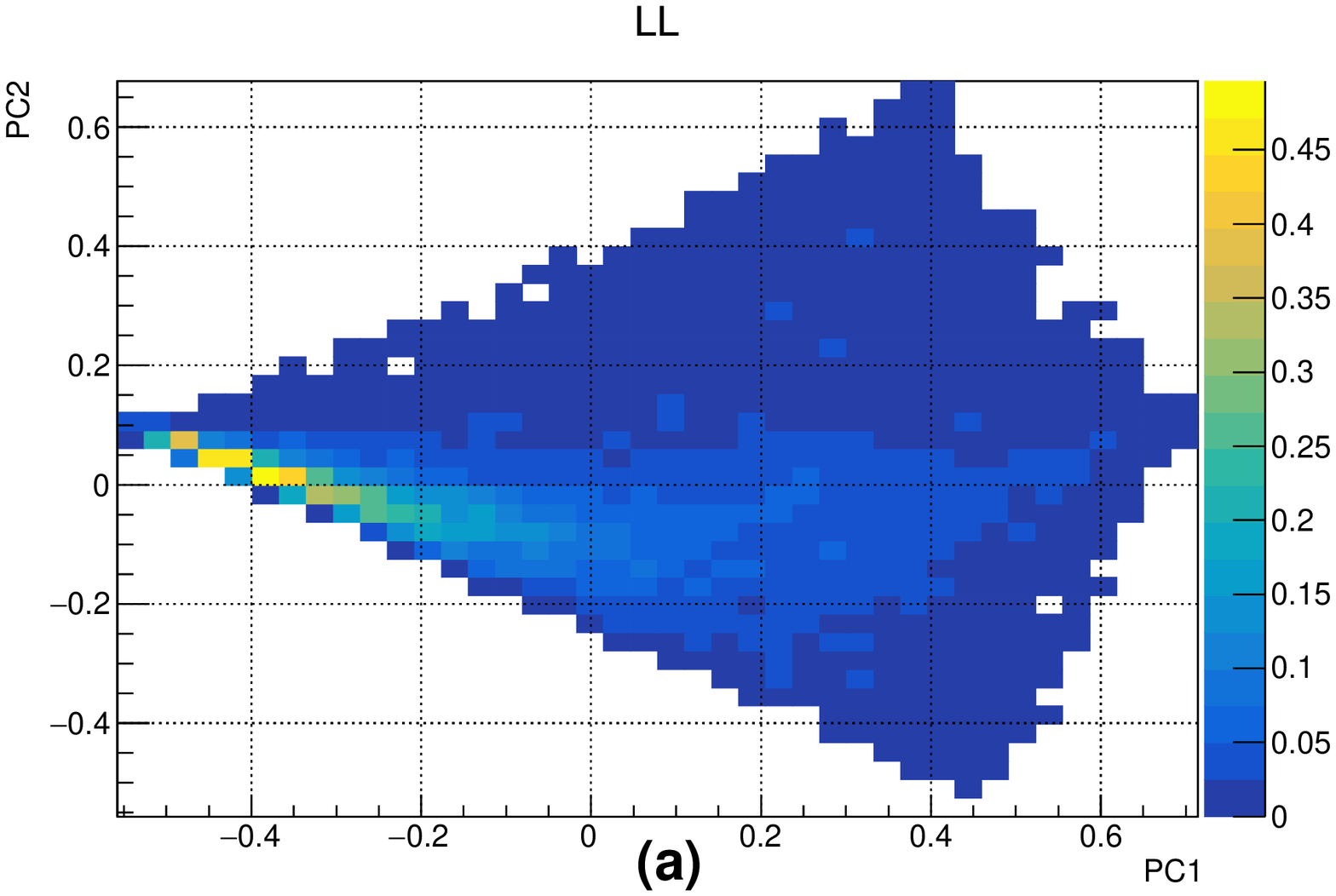}
  \includegraphics[width=0.4\textwidth]{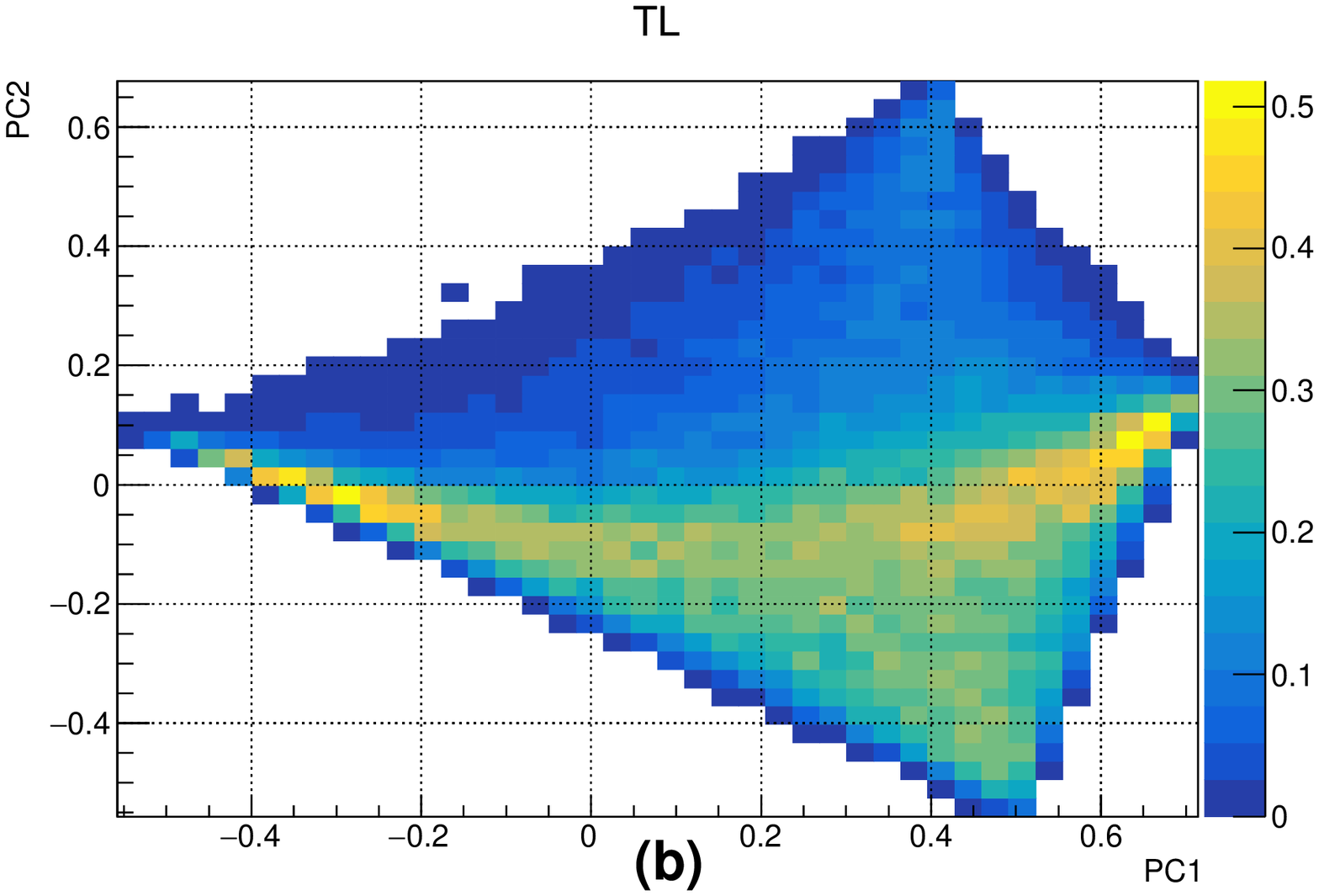}
  \includegraphics[width=0.4\textwidth]{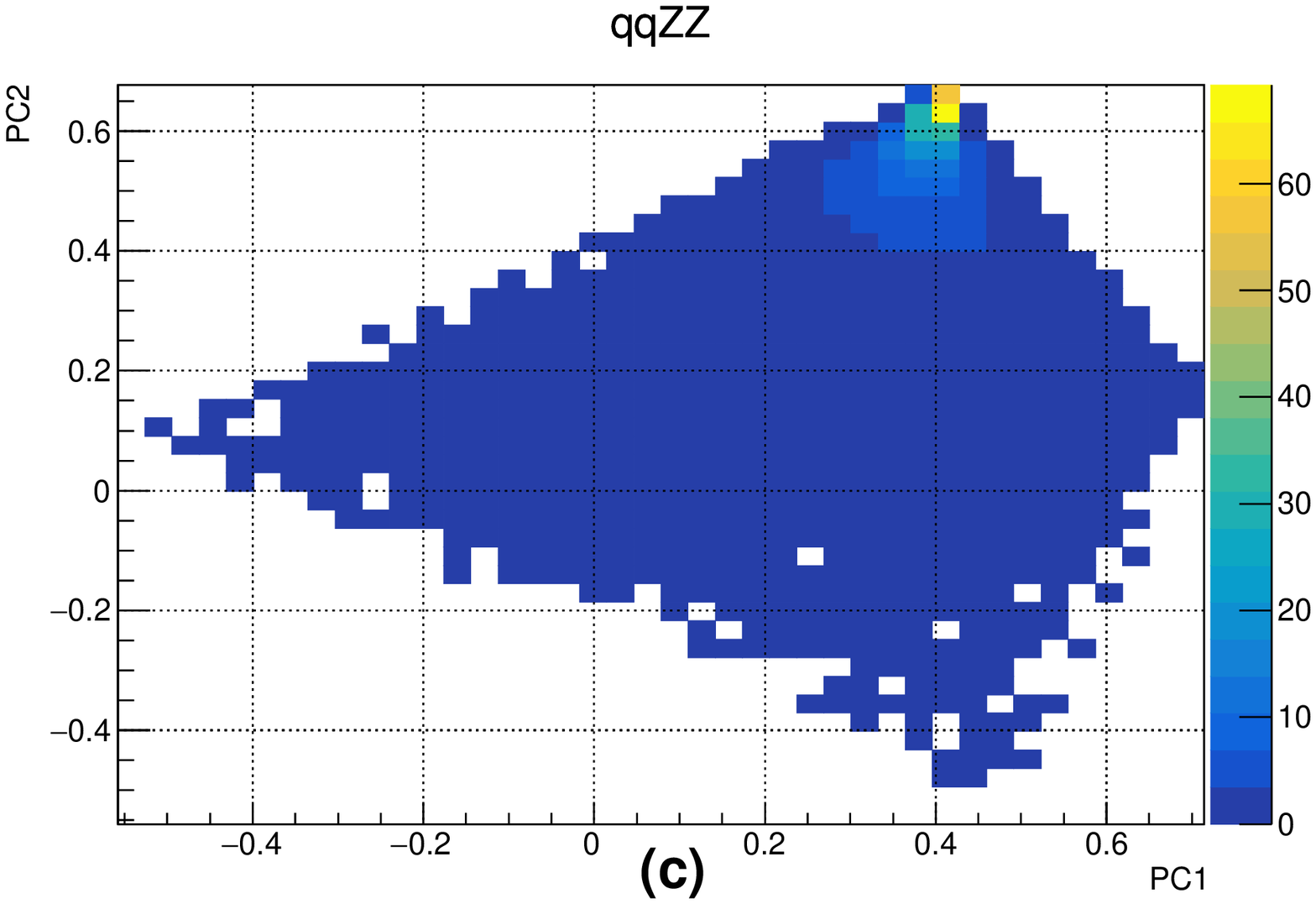}
    \caption{Two dimensional histograms based on PC1 and PC2. (a), (b), and (c) correspond to the LL, TL and qqZZ samples, respectively. Each histogram is normalized to the expected yield.}
    \label{fig:PCA_2D}
  \end{center}
\end{figure}

\begin{figure}[htbp]
  \begin{center}
  \includegraphics[width=0.4\textwidth]{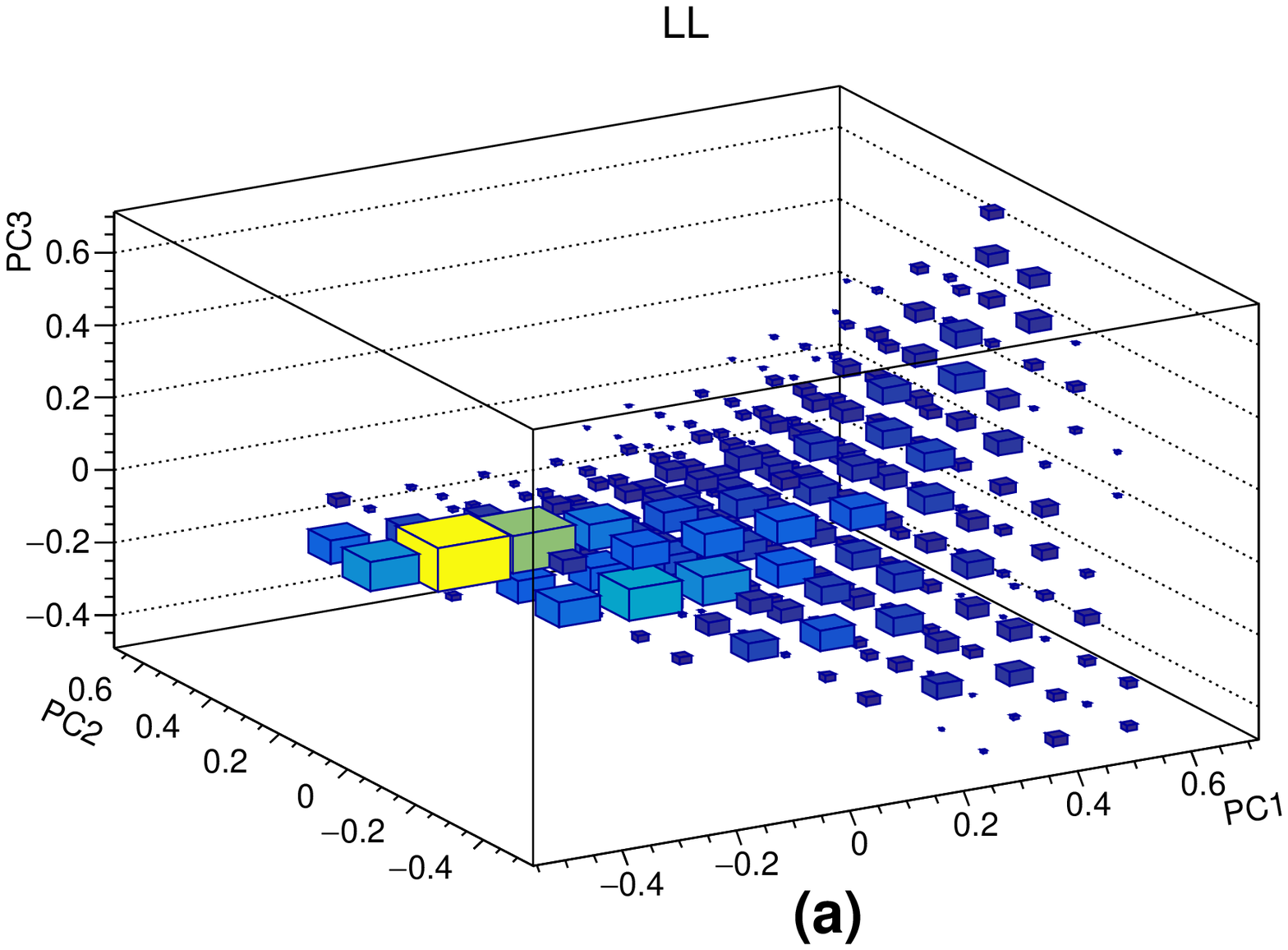}
  \includegraphics[width=0.4\textwidth]{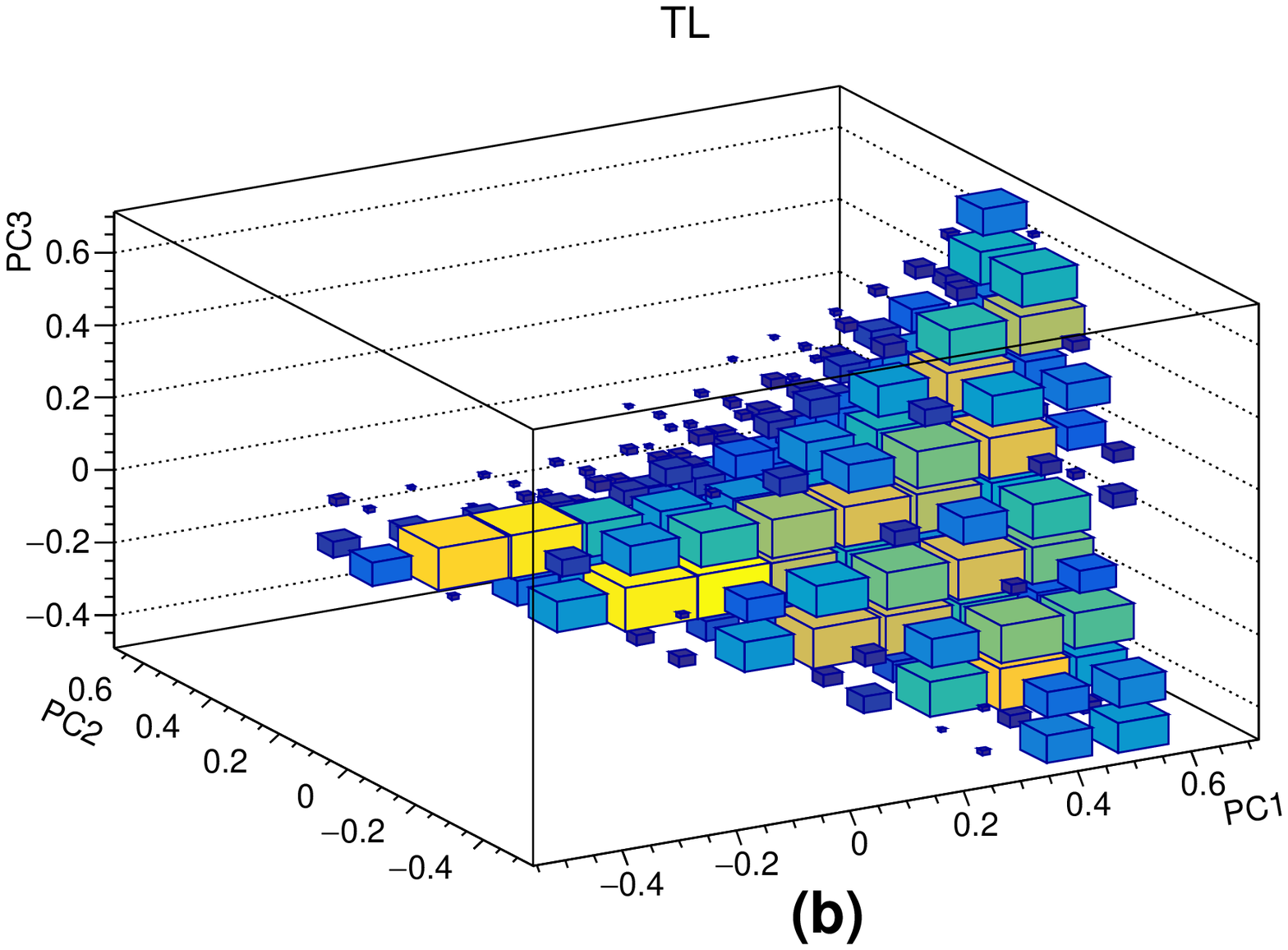}
  \includegraphics[width=0.4\textwidth]{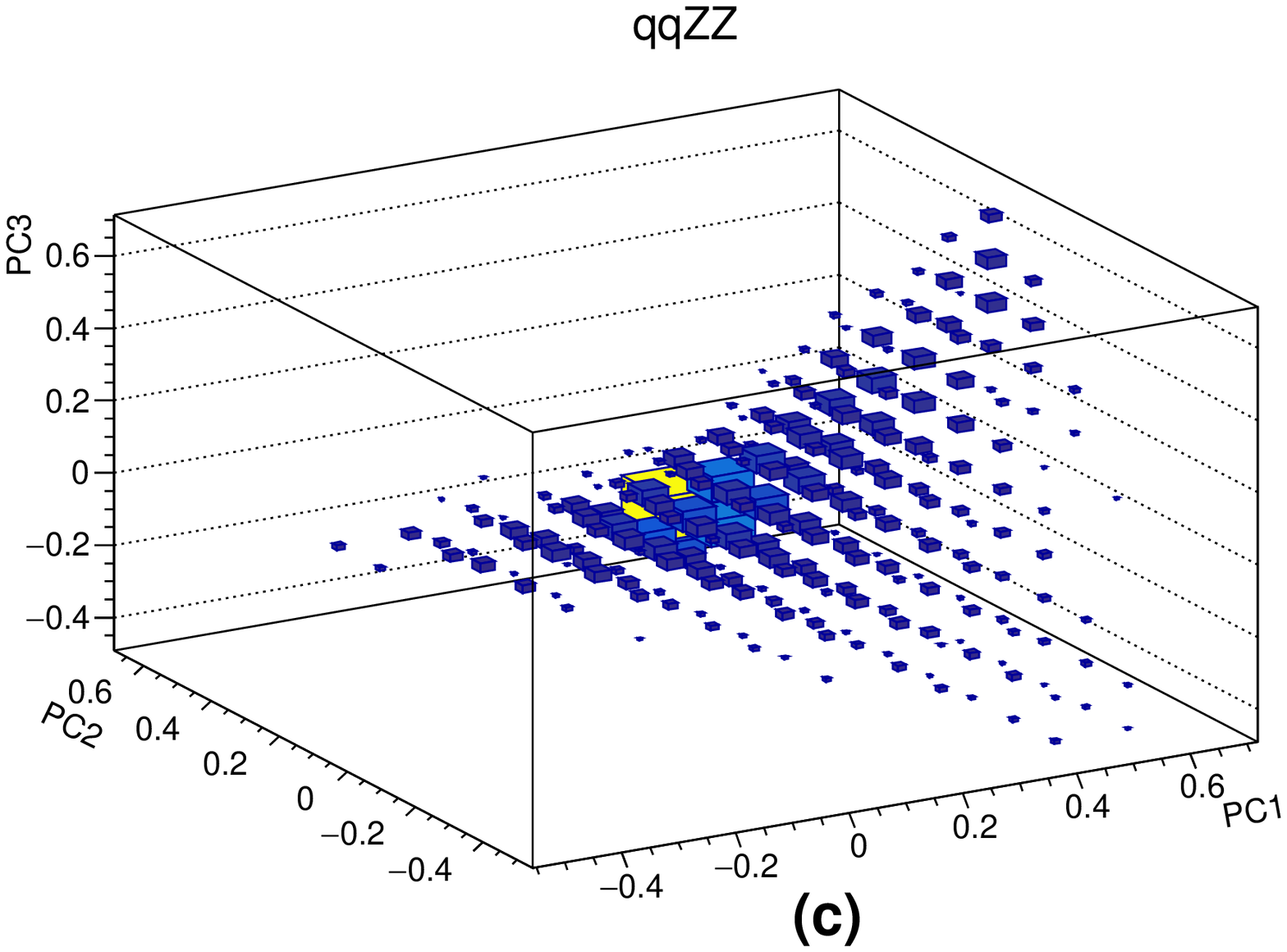}
    \caption{Three dimensional histograms based on PC1, PC2, and PC3. (a), (b), and (c) correspond to the LL, TL and qqZZ samples, respectively.}
    \label{fig:PCA_3D}
  \end{center}
\end{figure}

\begin{figure}[htbp]
  \begin{center}
  \includegraphics[width=0.4\textwidth]{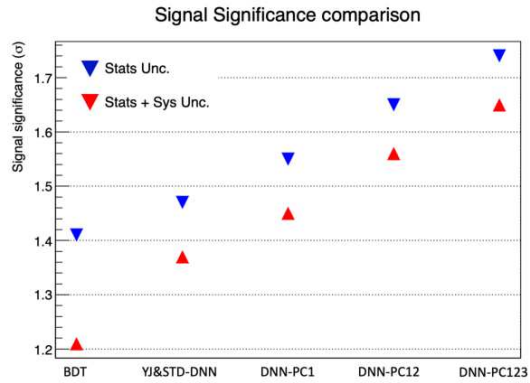}
      \caption{Expected signal significances obtained with the various algorithms that we have tested. The blue triangles were calculated with only statistical uncertainties, while the red triangles were calculated with both statistical and systematic uncertainties.}
    \label{fig:significance_plot}
  \end{center}
\end{figure}

We have investigated if the signal significance could be enhanced by using a combination of PC1, PC2, PC3.
Fig.~\ref{fig:PCA_2D} shows two dimensional histograms of PC1 and PC2 for the LL, TL and qqZZ samples.
Using the PC1 and PC2 two-dimensional distribution, a significance of 1.65$\sigma$ is obtained (1.57$\sigma$ when systematic uncertainties are considered).
The 3-dimensional distributions of PC1, PC2, and PC3 are shown in Fig.~\ref{fig:PCA_3D}, and a significance of 1.74$\sigma$ is achieved when they are used to extract the signal (1.66$\sigma$ when systematic uncertainties are considered).
This scenario has the largest signal significance among all of the models we have investigated. Fig.~\ref{fig:significance_plot} summarizes the expected significances obtained with various algorithms we have studied.

In summary, unitarity restoration or possible new physics can be probed by measuring longitudinally polarized VBS. And the VBS ZZ~$\rightarrow$ 4l channel, with its clean final state, is one of the most promising channels for this purpose.
We have investigated the performance of several Deep Neural Network structures in the task of extracting the longitudinally polarized component of ZZ scattering.
The DNN structure giving the best results was found to be a particle-based DNN~\cite{Lee:2018xtt} working on pre-processed data, that are further decorrelated with principle component analysis and used in a multi-dimensional fit to achieve separation of the different kind of backgrounds.
Using fast simulation, about 40\% improvement in the expected significance is found with respect to a previous study using a BDT.
Assuming an integrated luminosity at the HL-LHC of 3000 $fb^{-1}$ and considering all backgrounds, a significance of 1.7 standard deviations is expected, which would be increased to larger than 2 standard deviations when CMS and ATLAS measurements are combined. The technique developed in this article is also useful to other LHC analyses involving helicity fraction measurement. Moreover, as shown in Fig.~\ref{fig:ROC_pilot_curves} [left], longitudinally polarized VBS component can be sensitive to deviated Higgs couplings and also dimension-8 anomalous gaugue boson couplings, a more detailed studies to probe these new physics using machine learning is being examined, in referring to methods reported recently~\cite{Brehmer:2018eca,Brehmer:2018kdj}.

\pagebreak
\newpage
\acknowledgements
This work is supported in part by the National Natural Science Foundation of China, under Grants No. 11575005, by MOST under grant No. 2018YFA0403900, and COST Action CA16108. We thank the CNRS/IN2P3 and the France China Particle Physics Laboratory (FCPPL) for their support.

\bibliographystyle{ieeetr}
\bibliography{h}

\begin{thebibliography}{10}

\bibitem{Sirunyan:2017ret}
  A.~M.~Sirunyan {\it et al.} [CMS Collaboration],
  Phys.\ Rev.\ Lett.\  {\bf 120}, no. 8, 081801 (2018)
  doi:10.1103/PhysRevLett.120.081801
  [arXiv:1709.05822 [hep-ex]].

\bibitem{Aaboud:2019nmv} 
  M.~Aaboud {\it et al.} [ATLAS Collaboration],
  arXiv:1906.03203 [hep-ex].

\bibitem{Sirunyan:2019ksz} 
  A.~M.~Sirunyan {\it et al.} [CMS Collaboration],
  Phys.\ Lett.\ B {\bf 795}, 281 (2019)
  doi:10.1016/j.physletb.2019.05.042
  [arXiv:1901.04060 [hep-ex]].

\bibitem{ATLAS:2018ucv} 
  The ATLAS collaboration [ATLAS Collaboration],
  ATLAS-CONF-2018-033.

\bibitem{Khachatryan:2017jub} 
  V.~Khachatryan {\it et al.} [CMS Collaboration],
  Phys.\ Lett.\ B {\bf 770}, 380 (2017)
  doi:10.1016/j.physletb.2017.04.071
  [arXiv:1702.03025 [hep-ex]].

\bibitem{CMS:2019iuv} 
  CMS Collaboration [CMS Collaboration],
  CMS-PAS-SMP-18-007.

\bibitem{ATLAS-CONF-2019-033}
  ATLAS Collaboration [ATLAS Collaboration],
  ATLAS-CONF-2019-033

\bibitem{Sirunyan:2017fvv}
  A.~M.~Sirunyan {\it et al.} [CMS Collaboration],
  Phys.\ Lett.\ B {\bf 774}, 682 (2017)
  doi:10.1016/j.physletb.2017.10.020
  [arXiv:1708.02812 [hep-ex]].
\bibitem{Chang:2013aya} 
  J.~Chang, K.~Cheung, C.~T.~Lu and T.~C.~Yuan,
  Phys.\ Rev.\ D {\bf 87}, 093005 (2013)
  doi:10.1103/PhysRevD.87.093005
  [arXiv:1303.6335 [hep-ph]].

\bibitem{Lee:2018fxj} 
  S.~J.~Lee, M.~Park and Z.~Qian,
  Phys.\ Rev.\ D {\bf 100}, no. 1, 011702 (2019)
  doi:10.1103/PhysRevD.100.011702
  [arXiv:1812.02679 [hep-ph]].

\bibitem{Quigg:2018llo}
  C.~Quigg,
  Rev.\ Accel.\ Sci.\ Tech.\  {\bf 10} (2019) no.01,  3
  doi:10.1142/9789811209604\_0002, 10.1142/S1793626819300020
  [arXiv:1808.06036 [hep-ph]].

\bibitem{Brooijmans:2014eja} 
  G.~Brooijmans {\it et al.},
  arXiv:1405.1617 [hep-ph].

\bibitem{CMS:2018zxa} 
  CMS Collaboration [CMS Collaboration],
  CMS-PAS-FTR-18-005.

\bibitem{Lee:2018xtt}
  J.~Lee, N.~Chanon, A.~Levin, J.~Li, M.~Lu, Q.~Li and Y.~Mao,
  Phys.\ Rev.\ D {\bf 99}, no. 3, 033004 (2019)
  doi:10.1103/PhysRevD.99.033004
  [arXiv:1812.07591 [hep-ph]].

\bibitem{CMS:2018mbt}
  CMS Collaboration [CMS Collaboration],
  CMS-PAS-FTR-18-014.

\bibitem{Roe:2004na}
  B.~P.~Roe, H.~J.~Yang, J.~Zhu, Y.~Liu, I.~Stancu and G.~McGregor,
  Nucl.\ Instrum.\ Meth.\ A {\bf 543}, no. 2-3, 577 (2005)
  doi:10.1016/j.nima.2004.12.018
  [physics/0408124].

\bibitem{TMVA}
 Hocker, A. et al. TMVA - Toolkit for Multivariate Data Analysis. PoS ACAT, 040 (2007).

\bibitem{Keras}
 F.~Chollet et al., https://github.com/fchollet/keras

\bibitem{Tensorflow}
 Martín Abadi et al., TensorFlow: Large-scale machine learning on heterogeneous systems,
 2015. Software available from tensorflow.org.

\bibitem{10.1093/biomet/87.4.954}
  I.~Yeo, R.~A.~Johnson,
  Biometrika {\bf 87}, no. 4, 954-959 (2000)
  doi:10.1093/biomet/87.4.954

\bibitem{Alwall:2014hca}
  J.~Alwall {\it et al.},
  JHEP {\bf 1407}, 079 (2014)
  doi:10.1007/JHEP07(2014)079
  [arXiv:1405.0301 [hep-ph]].

\bibitem{Sjostrand:2003wg}
  T.~Sjostrand, L.~Lonnblad, S.~Mrenna and P.~Z.~Skands,
  hep-ph/0308153.

\bibitem{deFavereau:2013fsa}
  J.~de Favereau {\it et al.}  [DELPHES 3 Collaboration],
  JHEP {\bf 1402}, 057 (2014)
  [arXiv:1307.6346 [hep-ex]].

\bibitem{Searcy:2015apa}
  J.~Searcy, L.~Huang, M.~A.~Pleier and J.~Zhu,
  Phys.\ Rev.\ D {\bf 93}, no. 9, 094033 (2016)
  doi:10.1103/PhysRevD.93.094033
  [arXiv:1510.01691 [hep-ph]].


\bibitem{DBLP:journals/corr/HeZR015}
  K.~He, X.~Zhang, S.~Ren and J.~Sun
  CoRR {\bf abs/1502.01852} (2015)
  [arXiv:1502.01852 [cs]].

\bibitem[Kingma and Ba(2014)]{2014arXiv1412.6980K}Kingma, D.P., and Ba, J.: 2014, {\it arXiv e-prints} , arXiv:1412.6980.


\bibitem{10.2307/2984418}
  G.~E.~P.~Box and D.~R.~Cox, 
  Journal of the Royal Statistical Society. Series B (Methodological) {\bf 26}, no. 2, 211-252 (1964)
  
\bibitem{Cowan2011}
  G.~Cowan,
  The European Physical Journal C {\bf 71}, no. 2, 1554 (2011)
  doi:10.1140/epjc/s10052-011-1554-0

\bibitem{wiki:xxx}
  Wikipedia contributors,
  Principal component analysis --- {Wikipedia}{,} The Free Encyclopedia (2019)
  "\url{https://en.wikipedia.org/w/index.php?title=Principal_component_analysis&oldid=910232284}",

\bibitem{Brehmer:2018eca}
  J.~Brehmer, K.~Cranmer, G.~Louppe and J.~Pavez,
  Phys.\ Rev.\ D {\bf 98} (2018) no.5,  052004
  doi:10.1103/PhysRevD.98.052004
  [arXiv:1805.00020 [hep-ph]].

\bibitem{Brehmer:2018kdj}
  J.~Brehmer, K.~Cranmer, G.~Louppe and J.~Pavez,
  Phys.\ Rev.\ Lett.\  {\bf 121} (2018) no.11,  111801
  doi:10.1103/PhysRevLett.121.111801
  [arXiv:1805.00013 [hep-ph]].

\end{thebibliography}
\end{document}